\documentclass[floatfix,twocolumn,english,english,english,aps,prb,groupedaddress,showpacs]{revtex4}

\usepackage[dvips]{graphicx}
\usepackage[T1]{fontenc}
\usepackage[latin1]{inputenc}
\usepackage{babel}
\usepackage{dcolumn}
\usepackage{amsfonts}

\newcommand{\defect}[1]{\textsc{\textbf{#1}}}


\begin{document}
\title{Soliton effects in dangling-bond wires on Si(001)}
 
\author{C.F.Bird}

\email[]{c.bird@ucl.ac.uk}

\author{A.J.Fisher}

\email[]{andrew.fisher@ucl.ac.uk}
\altaffiliation[Also at ]{London Centre for Nanotechnology, Gordon St, 
London WC1E 6BT}
\author{D.R.Bowler}
 
\email[]{david.bowler@ucl.ac.uk}
\altaffiliation[Also at ]{London Centre for Nanotechnology, Gordon St, 
London WC1E 6BT}
\affiliation{Dept. of Physics and Astronomy, University College London, Gower
Street, London WC1E 6BT, U.K.}
 
\date{\today{}}

\begin{abstract}
Dangling bond wires on Si(001) are prototypical one dimensional wires,
which are expected to show polaronic and solitonic effects.  We
present electronic structure calculations,
using the tight binding model, of solitons in dangling-bond
wires, and demonstrate that these defects are stable in even-length
wires, although approximately 0.1\,eV higher in energy than a perfect
wire. We also note that in contrast to conjugated polymer systems, 
there are two types of soliton and that the
type of soliton has strong effects on the energetics of the bandgap
edges, with formation of intra-gap states between 0.1\,eV and 0.2\,eV
from the band edges. These intra-gap states are localised on the atoms
comprising the soliton.
\end{abstract}


\maketitle

\section{\label{sec:intro}Introduction}

There are many motivations to understand the transport properties of
materials in the extreme one-dimensional limit.  Some are
technological: the logical conclusion to the historic reduction in
size of electronic components would be active device elements, and
passive connections between them, that are of atomic scale.  There has
been much recent interest in structures that might act as atomic-scale
wires\cite{lyding94a} or as atomic- or molecular-scale switches.  Other
reasons relate to fundamental physics: transport in one dimension is
very different to higher dimensions, because the coupling of electrons
to other excitations, both of the lattice and of the electronic
system, is strong.  This can lead to instabilities such as the
formation of a Luttinger liquid\cite{haldane81a} or one of a
Peierls\cite{peierls_book_distor} (infinite wire) or Jahn-Teller\cite{jahn37a}
(finite wire) distortion, with correspondingly strong modifications to
transport properties.

One striking example of this type of behaviour is in conductive
polymer systems \cite{heeger88a}.  These exhibit an alternation of
double and single bonds which is at least partly of the classic
Peierls type (driven by electron-lattice coupling), although electron
correlation effects may also be important \cite{konig90}.  Carriers
introduced into these systems are localised, probably partly by
disorder but also to a significant degree by self-trapping by the
lattice \cite{landau33}.  There are two main classes of charged defect;
one is the polaron, which is similar in principle to excitations in
higher-dimensional systems.  Here the carrier is surrounded by a
region in which the atomic distortions serve to lower its own energy
(in this case, involving a reduction in the double-single bond
alternation).  The second type is the soliton; this can occur only in
`degenerate' systems, such as {\it trans}-polyacetylene (t-PA), where there
are two equivalent ground states of an infinitely long chain, in which
the double and single bonds are interchanged. The soliton involves a
mid-gap state associated with a `domain wall' between regions of
opposite bond alternation.  Although localized, both polarons and
solitons are highly mobile, and much is known about their effect on
transport in both the coherent and incoherent limits
\cite{heeger88a,ness99a,ness02a}.

In this paper, we examine an alternative pseudo-one-dimensional system
based on the ``dangling-bond wire''.  This system is formed on a
hydrogen terminated silicon (001) surface via STM-induced selective
desorption of hydrogen along the edge of a dimer row \cite{hitosugi99a}.
The atoms in such wires have been shown (both theoretically and
experimentally) to undergo a Peierls/Jahn-Teller type distortion,
giving rise to an alternating pattern of atomic positions with respect
to those of the passivated surface and producing the electronic
structure of a one-dimensional narrow-gap semiconductor
\cite{hitosugi99a}.  In contrast to the conducting polymers, this
distortion occurs predominantly perpendicular to the axis of the
`wire'; however, as in the conductive polymers, the resulting
structure is not expected to be ``static''. We recently predicted
\cite{bowler01b}, on the basis of tight-binding calculations, that
electrons or holes introduced into such wires will produce
self-trapped ``small polaron'' defects. Despite their localisation, we
have also shown that these defects are remarkably mobile near room
temperature and above\cite{todorovic02a}, raising the interesting
possibility of the transport of charge through such devices.

It is natural to ask whether there might be solitonic, as well as
polaronic, defects in dangling-bond wires (as in t-PA); after all,
these structures are degenerate (the total energy is invariant under
exchanging the ``up'' and ``down'' atoms).  This question is
intimately related to the effect of different types of boundary
condition; in the case of t-PA, double bonds are ``anchored'' at the
ends of the chain, so that the ground state of a neutral chain with
odd length contains a bond-alternation defect (soliton) at the centre.
The behaviour in t-PA is simple because the energy scales for the
binding of double bonds to the ends of the chain are much larger than
any involved in the formation of defects along it.

In this paper we explicitly focus on the behaviour of dangling-bond
wires of finite length, with a view to studying the formation of
solitons and the end effects. We find that the situation is more
complicated than in t-PA, because the energy scales of the end effects
and the defects are comparable to one another. This means that there
is much more complex behaviour when deliberate defects are
introduced into the alternating pattern. Accordingly we pay
particular attention to the stability of various possible defects and 
their contributions to the electronic structure of the wire. 
In light of the large number of possible initial configurations and necessarily
large system size ($>$400 atoms), we
chose to use a semi-empirical tight-binding approach\cite{goringe97a}
that was fast enough
to allow a comprehensive study of possible defects and had accurate
parametrisations available for the system under consideration.

\section{\label{sec:details}Technical Details}

\subsection{Basic configuration \& relaxation parameters}

A silicon slab was constructed out of 6 layers, parallel to the (001) plane.
Each layer had a 4 x 12 array of silicon atoms positioned at the appropriate
positions for a perfect ``diamond'' lattice, with the ``bottom'' layer terminated
with hydrogen atoms. This layer and the associated hydrogen atoms were not
allowed to move during relaxation and represented the bulk. 
The unterminated upper surface was allowed to reconstruct to form a pair of dimer
rows.
This (001) surface was then terminated with hydrogen atoms to produce a fully 
terminated slab that served as a basis for all the simulations.  This
unit cell contained 432 atoms.

Each slab was separated from its neighbours by a vacuum gap of 15\,\AA\ once
periodic boundary conditions were applied.
For the maximum length considered of eight dangling bonds, each wire was
isolated by four hydrogen terminated atoms from its virtual neighbours along the 
wire direction, while shorter wires had proportionately greater numbers of
hydrogen terminated atoms at each end.
All wires were isolated by a completely terminated dimer row from adjacent wires
once periodic boundaries were considered as this has been found to provide adequate
isolation \cite{healy01a}.

Wires were created by removing the required number of terminating hydrogen atoms
from the upper surface, along one side of a dimer row. 
The reference point for distortions
was the average position of  hydrogen terminated silicon atoms in the passivated
surface, labelled as the baseline or ``level'' position. Displacements normal
to the surface in the direction of the bulk were ``down'' and away from
the bulk were ``up''. Before relaxation, up atoms were displaced by an additional 0.3\,\AA\ compared 
to terminated atoms of the normal spacing and down atoms by 0.4\,\AA\ less than 
terminated atoms.

Previous work\cite{bird01a} had examined the effect of slightly perturbing
the starting positions to determine whether the final results were stable 
with respect to small distortions and found this was the case. Accordingly 
only one magnitude of starting displacement was considered in this study.

All relaxations and electronic structure calculations were performed using the
Oxford Order-N (OXON) tight-binding package\cite{goringe97a}. 
Structural relaxations were performed at a system temperature of 0 K until 
the maximum force per atom  reached the limit
of 0.01\,eV/\AA. The majority of simulations required in the order of 150 iterations
to relax. The Hamiltonian was solved using exact diagonalisation and periodic 
boundary conditions were applied to the simulation. Spin-polarisation was
not included in the calculations. The silicon parametrisation used was that of
Bowler \textit{et al}\cite{bowler98a}, which was specifically fitted
for the Si(001) surface and hydrogen on that surface.  It is worth
noting that this parameterisation did not fit to the conduction band,
so that absolute values should not be trusted, though generic
behaviour (such as shifts in levels) are likely to be correct.

Convergence of the total energy was tested with respect to the \textit{k}-point mesh.
An initial search was performed with a 3 atom wire to minimise calculation
time, and selected results replicated with an 8 atom wire. 
Table~\ref{tab:k_pt_conv} shows the results of the convergence calculations. Results 
were adequately converged (<0.01\,eV) using a Monkhorst-Pack\cite{monkhorst76a}
2x1x1 grid, and all simulations presented below used such a mesh. 

\begin{table}
\begin{ruledtabular}
\begin{tabular}{cdd}

\multicolumn{1}{c}{\text{Mesh Size}}&
\multicolumn{1}{c}{\text{3 atom chain}}&
\multicolumn{1}{c}{\text{8 atom chain}}\\

\hline

1x1x1&
-1963.7827&
-1948.6625\\

1x2x1&
-1963.7828&
-1948.6625\\

2x1x1&
-1964.1376&
-1949.0197\\

2x2x1&
-1964.1377&
-1949.0197\\

2x4x1&
-1964.1377&
---\\

4x1x1&
-1964.1364&
-1949.0183\\

4x2x1&
-1964.1365&
---\\

4x4x1&
-1964.1365&
---\\

4x4x4&
-1964.1365&
---\\

\end{tabular}
\end{ruledtabular}
\caption{\label{tab:k_pt_conv} Total energies (eV) calculated for 
        relaxed perfectly ordered 3 and 8 atom long chains at 
	0~K as a function of \textit{k}-point mesh size.}
\end{table}

\subsection{Configurations considered}

Dangling bond wires 4, 6 and 8 atoms long were considered. Some simulations of
odd length wires were performed, but these were not pursued for reasons discussed
below in Section~\ref{subsec:term}. Each wire was considered
initially with perfect structure, i.e.\ the initial starting positions were 
alternately up then down (or vice versa). Defects were then introduced
consisting of two adjacent up atoms or two adjacent down atoms at all possible 
positions in the wire with the remaining atoms alternating appropriately.

\begin{figure*}
  \includegraphics[width=\textwidth]{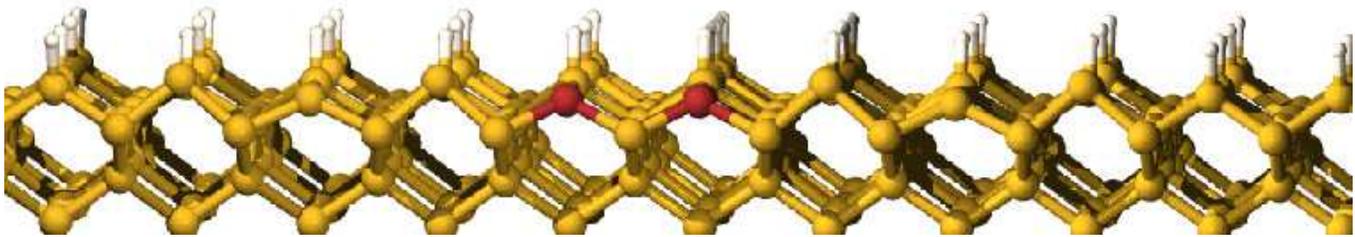}
  \caption{\label{fig:wire}A DB wire of length eight, with a DD
  soliton in the middle (marked with darker atoms).  The image shows
  only the top four layers of silicon, and limited terminating
  hydrogens (one on the left and two on the right).}
\end{figure*}

\section{\label{sec:results}Results}

\subsection{\label{subsec:term}End Effects}
We have previously tested the use of tight binding for infinite dangling bond
wires \cite{bowler01b,todorovic02a}, obtaining good agreement with experiment and
\textit{ab-initio} results. However, when we came to examine finite wires
with an odd number of sites, we found an anomaly: one site on the wire 
became ``level'' (i.e. at the same height as if it were hydrogenated) and
the rest of the wire behaved as a wire with an even number of sites, in 
contrast to \textit{ab-initio} results.  The local charge on the level
atom was zero, again in contrast to the rest of the wire (where down
atoms showed a charge deficit and the up atoms showed a charge
excess).  In this section, we explore the
reasons for this behaviour and investigate whether tight binding is a viable
technique for such systems.

The shortest, and simplest, finite wires of odd and even length for
which there is experimental data are those with three and four sites
respectively \cite{hitosugi99a} (the length two wire might be too
short, as it effectively consists of only end atoms). We have already
performed extensive investigations of this system using spin-polarised
density functional theory, and found good agreement with
experiment \cite{bird03a}. In particular, we found that
the length three wire has a rather small displacement pattern (with
the two end atoms 0.11\,\AA\ higher than the central atom) in close
agreement with experiment (where displacements of 0.15\,\AA\ are
measured), and that the length four wire can form alternating
up/down or down/up patterns, or an up/down/down/up pattern.

Our tight binding simulations of even length wires were in good
agreement both with the experimental results and the DFT simulations.
In order to better understand the peculiar behaviour for the odd
length wires, we used local charge neutrality (LCN) \cite{sutton88a}
(where the on-site energies of individual atoms are adjusted until
their net charge is zero) to remove any charge transfer effects in the
tight-binding simulations (it is worth noting that it is still
perfectly possible to obtain buckled dimers on the clean Si(001)
surface with a LCN condition---we will consider the effects on an even
length wire below).  The length three wire became almost flat under
this condition---though the two end atoms were 0.06\,\AA\ higher than
the central atom, in qualitative agreement with experiment and DFT
modelling.  We also modelled the length four wire with LCN, and found
little change---the alternating up/down pattern was still observed.
However, the LCN simulations were extremely hard to converge, and
did not show particularly improved results.
We conclude that, for odd length wires, there is spurious charge
transfer that makes modelling of these systems potentially inaccurate
in tight binding, and we will not consider them further\footnote{If two length three wires are
modelled on adjacent dimer rows, one shows an up/down/up pattern and
the other down/up/down, showing that, while tight binding allows
charge transfer too easily, it can model the basic physics of the
system correctly.}.  However, for even length wires there is no problem
and we will continue to model these using tight binding with
confidence.

\subsection{\label{subsec:solitons}Solitons in even length wires}

Simulations were performed for chain lengths of 4, 6 and 8 dangling-bond
systems. Initial configurations were chosen so as to include all possible
permutations of single position ordering defects, for both up and down
types. All relaxations successfully converged to the force limit. The
results are shown in detail in Table \ref{tab:even_wire}, but in summary both UU and DD soliton
defects could be formed in the relaxed wire. Compared to a perfectly
ordered wire, those containing solitons are approximately 0.1\,eV less stable.
No soliton defects formed on the end atoms and there was strong periodicity
in possible locations of the defects, as seen in the table.

\begin{table}
\begin{ruledtabular}
\begin{tabular}{cdddc}

Final&
\multicolumn{1}{c}{\text{E$_{\text{diff}}$}}&
\multicolumn{1}{c}{\text{VB Offset}}&
\multicolumn{1}{c}{\text{CB Offset}}&
Initial\\
\hline

\hline
%
%

udud&
\textemdash&
\textemdash&
\textemdash&
udud\\

dudu&
\textemdash&
\textemdash&
\textemdash&
dudu\\

udud&
\textemdash&
\textemdash&
\textemdash&
\defect{dd}ud\\

udud&
\textemdash&
\textemdash&
\textemdash&
ud\defect{uu}\\

u\defect{dd}u&
0.111&
-0.003&
-0.109&
u\defect{dd}u\\

u\defect{dd}u&
0.111&
-0.003&
-0.110&
\defect{uu}du\\

d\defect{uu}d&
0.102&
0.224&
0.012&
d\defect{uu}d\\

d\defect{uu}d&
0.101&
0.224&
0.012&
du\defect{dd}\\

\hline
%
%

dududu&
\textemdash&
\textemdash&
\textemdash&
dududu\\

ududud&
\textemdash&
\textemdash&
\textemdash&
ududud\\

ududud&
\textemdash&
\textemdash&
\textemdash&
\defect{dd}udud\\

ududud&
\textemdash&
\textemdash&
\textemdash&
udud\defect{uu}\\

u\defect{dd}udu&
0.114&
-0.003&
-0.106&
u\defect{dd}udu\\

u\defect{dd}udu&
0.114&
-0.002&
-0.106&
\defect{uu}dudu\\

udu\defect{dd}u&
0.114&
-0.002&
-0.106&
ud\defect{uu}du\\

udu\defect{dd}u&
0.114&
-0.002&
-0.106&
udu\defect{dd}u\\

d\defect{uu}dud&
0.104&
0.226&
0.007&
du\defect{dd}ud\\

d\defect{uu}dud&
0.104&
0.226&
0.006&
d\defect{uu}dud\\

dud\defect{uu}d&
0.104&
0.226&
0.007&
dudu\defect{dd}\\

dud\defect{uu}d&
0.104&
0.226&
0.007&
dud\defect{uu}d\\

\hline
%
%

dudududu&
\textemdash&
\textemdash&
\textemdash&
dudududu\\

udududud&
\textemdash&
\textemdash&
\textemdash&
udududud\\

dudududu&
\textemdash&
\textemdash&
\textemdash&
dududu\defect{dd}\\

dudududu&
\textemdash&
\textemdash&
\textemdash&
\defect{uu}dududu\\

u\defect{dd}ududu&
0.115&
0.000&
-0.102&
u\defect{dd}ududu\\

u\defect{dd}ududu&
0.115&
0.000&
-0.103&
ud\defect{uu}dudu\\

udu\defect{dd}udu&
0.117&
-0.001&
-0.105&
udu\defect{dd}udu\\

udu\defect{dd}udu&
0.117&
0.000&
-0.105&
udud\defect{uu}du\\

ududu\defect{dd}u&
0.115&
0.000&
-0.102&
ududu\defect{dd}u\\

ududu\defect{dd}u&
0.115&
0.001&
-0.102&
ududud\defect{uu}\\

d\defect{uu}dudud&
0.104&
0.228&
0.003&
\defect{dd}ududud\\

d\defect{uu}dudud&
0.105&
0.227&
0.004&
d\defect{uu}dudud\\

dud\defect{uu}dud&
0.106&
0.230&
0.009&
du\defect{dd}udud\\

dud\defect{uu}dud&
0.106&
0.230&
0.009&
dud\defect{uu}dud\\

dudud\defect{uu}d&
0.104&
0.227&
0.003&
dudu\defect{dd}ud\\

dudud\defect{uu}d&
0.104&
0.228&
0.003&
dudud\defect{uu}d\\

\end{tabular}
\end{ruledtabular}
\caption{\label{tab:even_wire}Final structure, energy change, valence
	and conduction band offsets and starting structure for soliton-containing even-length wires. %
	E$_{\text{diff}}$ represents difference in total energy relative to a perfect wire. %
	All energy differences are in electron-volts. Soliton defects are shown in bold capitals.
	VB stands for valence band, CB for conduction band.}
\end{table}

As well as showing the final, relaxed configuration of the wires (with
up atoms notated ``u'', down atoms ``d'' and atoms associated with a
soliton in bold capitals) and various energies,
Table~\ref{tab:even_wire} shows the initial configuration for each
system.  It is important to note that several different initial
configurations led to the same final configuration (for reasons
discussed below).

The effects of the ends of the wire are rather small (only a few meV)
but seem to exert a small attractive force on the solitons.  More
interesting is the apparent instability of solitons of both kinds
exactly at the ends of the wire (and of certain configurations with
the solitons in the middle of the wire).  For instance, the third and
fourth lines of table~\ref{tab:even_wire} show that the initial
configurations DDud and UUdu relaxed back to perfect wires, while the
eighth line shows that duDD relaxed back to dUUd.  This pattern can be
easily understood in terms of charge balance: an up atom is associated
with an excess of charge, while a down atom is associated with a
deficit of charge.  The configurations which were unstable (which
includes all systems with a single soliton at the end of the wire) had
different numbers of up and down atoms, which would lead to charge
imbalance, so they changed to a stable configuration with equal
numbers of up and down atoms, either removing
the soliton entirely or changing it.

\begin{table}
\begin{ruledtabular}
\begin{tabular}{ccdddddd}

Length&
Defect&
\multicolumn{1}{c}{\text{VB Edge}}&
\multicolumn{1}{c}{\text{CB Edge}}&
\multicolumn{1}{c}{\text{Bandgap}}\\
\hline

4&
Perfect&
-7.66&
-6.35&
1.31\\

&
\defect{dd}&
-7.66&
-6.46&
1.20\\

&
\defect{uu}&
-7.44&
-6.33&
1.11\\

\hline

6&
Perfect&
-7.66&
-6.36&
1.30\\

&
\defect{dd}&
-7.66&
-6.46&
1.20\\

&
\defect{uu}&
-7.43&
-6.35&
1.08\\

\hline

8&
Perfect&
-7.66&
-6.36&
1.30\\

&
\defect{dd}&
-7.66&
-6.46&
1.20\\

&
\defect{uu}&
-7.43&
-6.35&
1.08\\

\end{tabular}
\end{ruledtabular}
\caption{\label{tab:bandgap}
	Values of the bandgap in even-length wires containing various defects.
	VB stands for Valence Band, CB for Conduction Band, all energies in eV.}
\end{table}

\begin{figure}
{\centering \resizebox*{1\columnwidth}{!}{\includegraphics{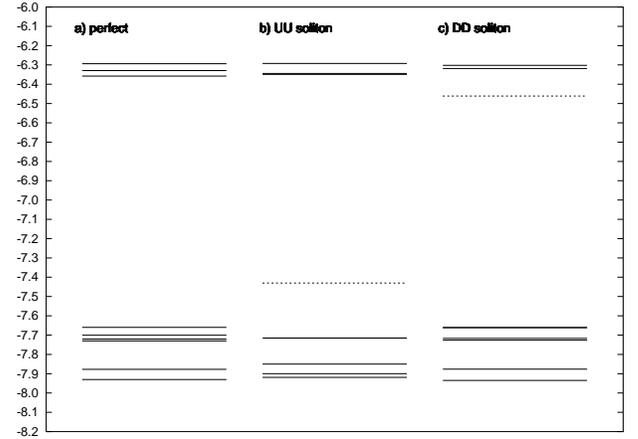}} \par}

\caption{\label{fig:band_edges}Energies (eV) of states adjacent to the band gap
for a) perfectly ordered 8 atom wire, b) 8 atom wire with UU defect, c) 8 atom wire
containing DD defect.}
\end{figure}

Looking at the electronic structure of relaxed wires containing
solitons, the presence of a soliton reduced the bandgap. The general
pattern is that a DD type defect lowered the bottom of the conduction
band, whilst an UU type defect raised the top of the valence band. The
values for different length wires are shown in table
\ref{tab:bandgap}.  If the soliton was of the DD type, the bottom of
the conduction band was lowered by 0.1\,eV, whilst the top of the
valence band was raised by 0.22\,eV by the presence of an UU
defect. Both these effects were due to the introduction of a intra-gap
state, as illustrated in Figure~\ref{fig:band_edges}. We recall 
that the system was not parameterised for the conduction band and
accordingly absolute energies relating to it should be treated with caution.

A soliton is generally associated with a mid-gap state, which is seen
here in the change of the valence and conduction bands.  These effects
can be understood in terms of the hybridisation of the silicon making
up the surface: in ideal, bulk positions the atoms are sp$^3$
hybridised, but reconstruction into dimers at the surface pulls the
atoms away from this state.  When Jahn-Teller distorted, either during
buckling of dimers or forming a finite length DB wire as here, the
atoms displaced up move closer to sp$^3$ hybridisation and gain
charge (tending to a lone pair) while the atoms displaced down move
closer to sp$^2$ hybridisation with a lone p-orbital and lose charge
(tending to an empty dangling bond).  Thus the up atoms are associated
with filled states at the top of the valence band, while the down
atoms are associated with empty states at the bottom of the conduction
band.  The up and down displacements affect the underlying substrate
(in particular the second and third layer atoms) in such a way that
alternating up and down displacements (either along or across a dimer
row) are energetically favourable.  When this is interfered with, as in
a soliton, the extent of the relaxation towards sp$^3$ or sp$^2$+p is
reduced, and the band edges are affected.  This is in marked contrast
to conjugated polymers, as will be discussed below.

\begin{figure}
{\centering \resizebox*{1\columnwidth}{!}{\includegraphics{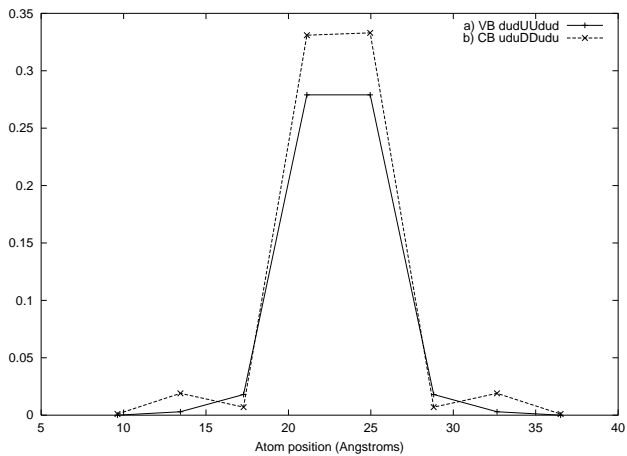}}\par}
\caption{\label{fig:localcharge}Effects of soliton formation on
  the electronic structure of finite DB wires.  We plot contributions to the charge density
  for specific bands from atoms most strongly affected as a function of distance along the wire;
  a) highest occupied electronic state for a central UU soliton, b)
  lowest unoccupied electronic state for a central DD soliton.}

\end{figure}

The solitons are broadly localised as expected, which is shown in
Figure~\ref{fig:localcharge}.  This is similar to the polaron seen in
the same system\cite{bowler01b,todorovic02a} and indicates that the
soliton is similarly weakly coupled to the bulk.  The overall
behaviour and form of the soliton in the DB wire is very different 
to the conjugated polymer system. In particular, the soliton
is associated with a pair of atoms and has two forms, while in
conjugated polymers it is associated with the bond lengths near a
single atom, and has only one form.

\section{\label{sec:discuss}Discussion and Conclusions}

Although the dangling-bond system is too reactive to be a component
of nanoscale circuitry, it does serve both as a potential model for
the study of one-dimensional conductors and as a stepping stone to
realisable useful systems, for example as a template for the deposition
of metal atoms in linear features. As such it is important that systems
can be modelled accurately and quickly. In this paper, we have demonstrated
that tight-binding simulations are suitable for even-length wires and
used such techniques to explore the stability of ordering defects in the
Jahn-Teller alternating displacement seen in perfect dangling bond wires.
We note that modelling of odd-length wires is complicated by the development of
``level'' atoms in the relaxed system. This appears to be related to
difficulties in adequately modelling unpaired electrons using the
tight-binding formalism; we have modelled these systems using
spin-polarised DFT, reported elsewhere\cite{bird03a}.

In the even length system, we demonstrate that ``soliton'' type defects, where
alternation between up and down displacements is interrupted, are
only $\sim 0.1$\,eV less stable than the perfect wire, but are associated
with strong effects on the edges on the valence and conduction bands,
as shown in Figure \ref{fig:band_edges}, as a consequence of changes in
hybridisation state.
The presence of an ``UU'' defect leads to the formation of an isolated
state 0.22\,eV above the original valence band. A ``DD'' defect is associated
with a state 0.1\,eV below the original conduction band, although the
conduction band energies should be treated with caution due to the nature
of the parametrisation. We also demonstrate that these states contain
large contributions from the atoms associated with the defect i.e localisation.

\begin{acknowledgments}
We thank the UK Engineering and Physical Sciences Research Council
for funding (CFB) and the Royal Society for a University Research
Fellowship (DRB).
\end{acknowledgments}

\bibliographystyle{apsrev}
\bibliography{strings,general}

\end{document}